# Strategies for deliveries of anti-cancer drugs from perspectives of a measurement theory and an adsorption theory

Ken-ichi Amano [a][*] and Takumi Otake [a]

[a] *Faculty of Agriculture, Meijo University, Nagoya, Aichi 468-8502 , Japan .*

[*] Correspondence author: K. Amano (amanok@meijo-u.ac.jp)

**ABSTRACT**

In this letter, we present anti-cancer drug delivery strategies using knowledges obtained from our recent studies. We have conducted inverse analyses of "density distributions of colloidal particles near a focused surface" and "pair potentials between the surface and the colloidal particle" using data measured by optical tweezers (OT) and atomic force microscopy (AFM). Non-additive Asakura-Oosawa (NAO) theory and a lattice theory in statistical mechanics of simple polymers have been also our research topics. Summarizing the knowledges, we propose two strategies to increase the delivery rate of capsule shaped anti-cancer drugs to cancer cells. We consider that enhanced repulsion between the normal cell and the drug accelerates the attraction between the cancer cell and the drug, which can be named enhanced repulsion and accelerated adsorption (ERAA) effect. To realize the ERAA effect, we propose a supporting method for measuring the interactions using OT and AFM. In the second strategy, dose of water-soluble polymers is considered to realize adsorptions of the drugs and cancer cell derived exosomes onto the cancer cells, which we call non-specific and selective adsorption (NSSA) effect. In the main text, we explain the NSSA effect using the NAO theory. Moreover, we explain that structural stabilities of normal dispersed proteins around the cancer cells are not largely destroyed by the dosed polymers from the viewpoint of the lattice theory.

**MAIN TEXT**

Based on the drug delivery system (DDS) and the enhanced permeability and retention (EPR) effect [1], capsule shaped anticancer drugs (colloidal particles) are designed to efficiently reach to the cancer cells. The targeted delivery of the drug leads to reduce of the side effects on the normal cells. It can improve the quality of life of the cancer patient. However, it has been reported that the median value in the delivery rates of the drugs to the cancer cells is only 0.7% [1]. Despite more than ten years of the research, it seems that the delivery rate has not drastically improved, yet. Hence, from a different viewpoint, it can be expected that there is a lot of room for improvement in the delivery rate. Therefore, we challenged to construct strategies to improve the delivery rate, where we make use of our measurement and adsorption theories. The strategies are proposed from viewpoints of the analytical chemistry and the theoretical physical chemistry.

In our past study, we have proposed the inverse analysis theories for obtaining the density distribution of dispersed colloidal particles near a surface [2-4] and the pair potential between a surface and a colloidal particle [5,6]. In the inverse analysis theories, the main input datum is an experimental data measured by OT [7] or AFM [8-10], and the theories are constructed based on statistical mechanics of simple liquids [11]. We have also studied the adsorption behaviors of colloidal particles (they have cores) in aqueous polymer solutions by originally deriving the NAO theory [4,12]. Moreover, we have also studied the structural stability of a folded protein in both an aqueous solution and an aqueous polymer solution using the lattice theory in statistical mechanics of simple polymers. Conclusion from the lattice theory is that the structural stability does not change in the aqueous polymer solution, even though the system is relatively crowded than the aqueous solution. The supporting result has been also obtained from our original spectroscopic measurements (not shown). We note that it is assumed in the lattice theory that the polymers do not have specific reactivity and interaction with water molecules and proteins. Summarizing such findings, we constructed two strategies for improving the delivery rate of the capsule shaped anti-cancer drugs.

The first strategy is to obtain not only the pair potential between the cancer cell and the drug but also the pair potential between the normal cell and the drug by using our inverse analysis theory [5,6]. If the pair potential between the normal cell and the drug is repulsive dominant, it is possible to further improve the delivery rate of drugs to the cancer cells (see Figure 1). We call this enhanced repulsion and accelerated adsorption (ERAA) effect.

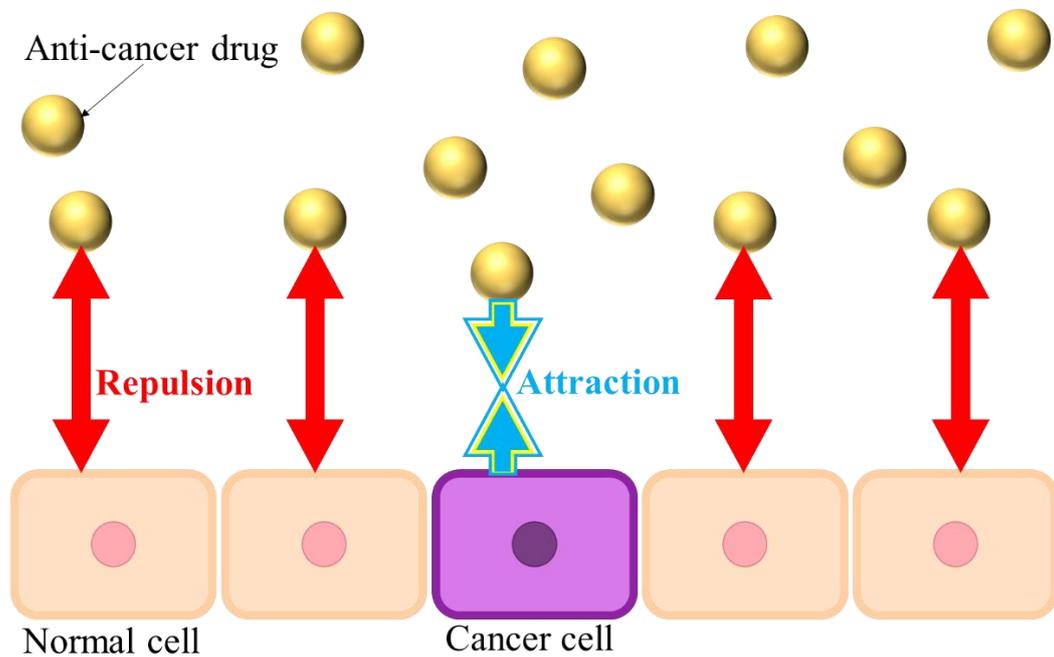

Figure 1: Illustration of the ERAA effect. Using both OT and the inverse analysis theory, the interaction between the normal cell and the drug is measured. It can help the development of surface modification of the capsule shaped anti-cancer drug. As the repulsion between the normal cell and the drug becomes stronger, the probability that the drug reaches to the cancer cell is increased. Hence, indirectly and relatively, the attraction between the cancer cell and the drug is increased. This concept is important, because the most of the cells in a cancer patient are not cancer cells, but normal cells.

Many of the studies on the delivery ratio have been conducted by focusing attractive interaction between the cancer cell and the drug following the EPR theory. Hence, as the counter strategy, we focus on the repulsion between the normal cell and the drug. This is because, most of the cells in the cancer patient are not the cancer cells but the normal cells. We would like to support developments of the drugs with our inverse analysis theories. Results obtained from the inverse analysis theories will be a supporting information for the developments.

The repulsion between the normal cell and the drug may also be evaluated by using a device for intermolecular interaction analysis [13]. With the device, a reaction rate constant between a substrate modified with an antibody (antigen) and an antigen (antibody) dissolved in a solution can be obtained. Although the degree of adsorption can be evaluated from the ratio, it is difficult to analyze the strength of the repulsive force. In addition, the analysis cannot measure the interaction between the actual cell surface and the drug. Therefore, we believe that our method, which uses LT (AFM) and the inverse analysis theory, can obtains more practical data.

The second strategy is to dose water-soluble polymers that do not have specific interactions with the cancer cells and the various substances around the cancer cells. Adding the polymers, both the drugs and exosomes derived from the cancer cells are accumulated on the surfaces of the cancer cells (see Figure 2), which is the effect predicted by the NAO theory. We mention the effect as non-specific selective adsorption (NSSA) effect. However, the polymers have the risk of destroying structures of the dispersed proteins around the cancer cells. Fortunately, the risk is considered to be avoided from viewpoints of the lattice theory and spectroscopic measurements (not shown).

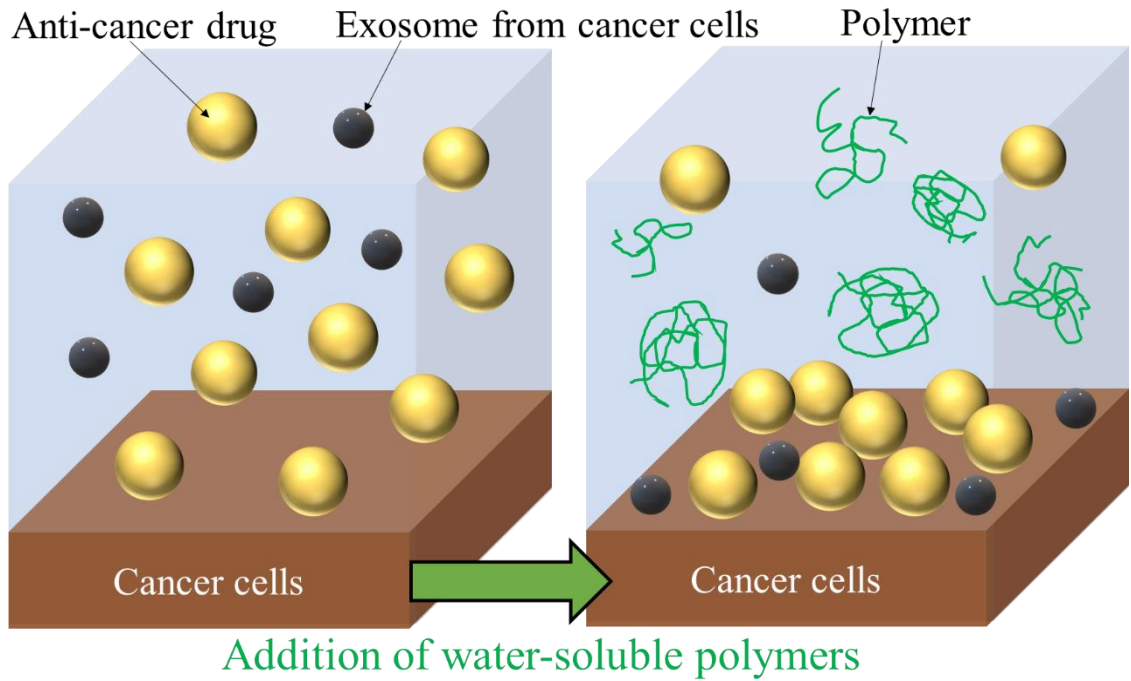

Figure 2: Illustration of the NSSA effect. Both capsule shaped anti-cancer drugs and the exosomes derived from the cancer cells are accumulated on surfaces of the cancer cells by dose of the water-soluble polymers, where the polymers do not have specific interactions. Mechanism of the accumulation can be explained using the NAO theory. In addition, structural stabilities of the normal dispersed proteins (not drawn) in the environment cannot be largely destroyed by dose of the polymers. It can be explained using the lattice theory.

To explain the NSSA effect in more detail, we explain the NAO effect [4,12]. The theory predicts degrees of adsorptions of particles with cores and particles without cores near a surface. For example, we consider an aqueous system that contains the anti-cancer drugs, the exosomes derived from the cancer cells, the water-soluble polymers, and the cancer cell surface. In the NAO theory, they are modeled as the spherical particles with cores, the spherical particles with cores, the spherical particles without cores, and the flat surface, respectively. In that case, the anti-cancer drugs and exosomes preferentially adsorb to the surface of the cancer cell, while the polymers do not preferentially adsorb to the surface compared with the drugs and the exosomes. That is, both the drugs and the exosomes can be selectively adsorbed to the surface by adding the polymers. In a simple form, the NSSA effect can be expressed by using a set of equations [4,12]:

$$g_{\text{CA,con}} = \exp(4\varphi_A + 4\varphi_E + 4\varphi_P), \tag{1}$$

$$g_{\text{CE,con}} = \exp(4\varphi_A + 4\varphi_E + 4\varphi_P), \tag{2}$$

$$g_{\text{CP,con}} = \exp(4\varphi_A + 4\varphi_E), \tag{3}$$

where the subscripts C, A, E, and P represent the cancer cell, anti-cancer drug, exosome, and polymer, respectively. The subscript con represents the contact point between the two substances. $g_{Ci,\text{con}}$ is the normalized number density of the species $i$ at the contact point with the surface of the cancer cell. $\varphi_i$ is the volume fraction of the species $i$ in the system. In equations (1)-(3), we approximate that the diameters of the anti-cancer drug, exosome, and polymer are the same. The anti-cancer drug and exosome are modeled as the spherical particles with rigid cores. The polymer is also spherical particle but it does not have the core, and hence the pair potential between the polymers is always zero. Equations (1)-(3) predict the NSSA effect. By the way, there is a concern that useful proteins around the cancer cells may also be adsorbed onto the cell surface unnecessarily by the addition of the polymers. However, if the NSSA effect is high enough, the concern can be avoided.

Next, we discuss a probable risk existing in the NSSA effect. The addition of the polymers may induce the adsorptions of the anti-cancer drugs and the exosomes, but we do not know whether useful proteins dispersed around the cancer cell are denatured or not by the added polymers. We considered the concern using the lattice model, and we predict that the addition of the polymers does not destroy structural stabilities of the proteins, if the polymers do not have specific reactivities and interactions with the proteins. Here, we focus on the amount of change in free energy ($\varDelta G$) that accompanies protein denaturation in a solution. $\varDelta G$

can be expressed by using the amount of change in enthalpy ($\Delta H$), the amount of change in entropy ($\Delta S$), and the absolute temperature ($T$) as follows: $\Delta G = \Delta H - T \Delta S$. To focus on the physical essence and for simplicity, we treat only $\Delta S$ afterward. In the lattice theory, $\Delta H$ is a term that is directly related to the interaction energy between two units (i.e., water unit-water unit, water unit-protein unit, water unit-polymer unit, protein unit-protein unit, protein unit-polymer unit, and polymer unit-polymer unit). When the attraction between the protein units is relatively strong, the protein's native structure is relatively stable. When the attraction between the protein unit and the polymer unit is relatively strong, denaturation or surface denaturation may occur in the protein structure due to addition of the polymers. Since the effects originate from the interaction parts can be predicted somewhat straightforwardly, we abbreviate $\Delta H$ in this discussion. Similar theoretical approaches are often practiced in a field of theoretical biophysics. In the lattice theory below, the conformational and mixing entropies are considered. To express the system entropy (S), we use the Boltzmann formula,

$$S = k_\text{B} \ln W \tag{4}$$

where $k_\text{B}$ is the Boltzmann constant and $W$ is the number of microscopic states. To calculate the entropy of a mixed system containing the waters, proteins, and polymers, we need to estimate the number of microscopic states. Since the estimation method for a binary system of waters and polymers in the reference [14], we extended the method to a ternary system of waters, proteins, and polymers. The number of microscopic states can be written as

$$W = \frac{1}{N_1! N_2! \sigma_1^{N_1} \sigma_2^{N_2}} \prod_{j=0}^{N_1-1} V_{j+1} \prod_{k=0}^{N_2-1} U_{k+1}, \tag{5}$$

where $V_{j+1}$ and $U_{k+1}$ are

$$V_{j+1} = z_1(z_1-1)^{n_1-2}\left(1-\frac{jn_1}{\Omega}\right)^{n_1-1}(\Omega-jn_1), \tag{6}$$

$$U_{k+1} = z_2(z_2-1)^{n_2-2}\left(1-\frac{N_1 n_1}{\Omega}-\frac{kn_2}{\Omega}\right)^{n_2-1}(\Omega-N_1 n_1-kn_2). \tag{7}$$

The subscripts 0, 1, and 2 represent the water, polymer, and protein, respectively. $N_0$, $N_1$ and $N_2$ are the numbers of waters, polymers, and proteins, respectively. $\sigma_1$ and $\sigma_2$ are the symmetric numbers, $z_1$ and $z_2$ are the numbers of nearest neighbor lattice points (arrangement numbers), $n_1$ and $n_2$ are the degrees of the polymerizations. In the lattice theory following rules are implicitly implemented. The proteins cannot take positions that have already been occupied by the polymers. Sizes of the unit structures are the same for the water, protein, and polymers. The protein is made by polymerizing the averaged amino acids, although various types of amino acids exist in the sequence of the real protein. The polymers do not have branches in their sequences. By the way, $\sigma_1$ and $\sigma_2$ written above are the symmetric numbers, which are generally 2 for a symmetric polymer and 1 for an asymmetric polymer. For the polymer, we set $\sigma_1 = 2$. For the protein, we set $\sigma_2 = 1$, because the real protein is asymmetric. (However, in the lattice model, the protein model is symmetric.) $\Omega$ is the total number of lattice cells in the system. Using the equations above and Stirling's formula etc., $\ln W$ is simplified as follows

$$\ln W = \frac{\Omega \phi_1}{n_1}\ln\left(\frac{n_1 \delta_{\max,1}}{\phi_1 \sigma_1 e^{n_1-1}}\right) + \frac{\Omega \phi_2}{n_2}\ln\left(\frac{n_2 \delta_{\max,2}}{\phi_2 \sigma_2 e^{n_2-1}}\right) - \Omega((-\phi_1-\phi_2)\ln(1-\phi_1-\phi_2)), \tag{8}$$

where $\varphi_1$ and $\varphi_2$ are the volume fractions of the polymers and proteins, respectively. $\delta_{\max,1}$ and $\delta_{\max,2}$ are the maximum degrees of conformational flexibility, which are given by

$$\delta_{\max,1} = z_1(z_1-1)^{n_1-2}, \tag{9}$$

$$\delta_{\max,2} = z_2(z_2-1)^{n_2-2}. \tag{10}$$

Hence, the amount of changes in the structural entropies due to the protein denaturation in a system with the polymer and without the polymer, $\Delta S_\text{N}$ and $\Delta S_\text{M}$ respectively, can be calculated using equation (8) (the subscripts N and M represent the normal and mixed conditions). First, we calculate $\Delta S_\text{N}$ (see Figure 3 upper). For example, the protein's conformational flexibility in the native structure can be assumed to be 1 in a rough approximation. On the other hand, the protein's conformational flexibility in the denatured structure can be expressed by using equation (10). Taking advantage of the facts that $\varphi_1$ is zero in the normal system and $\varphi_2$ does not change through the denaturation, $\Delta S_\text{N}$ can be expressed as

$$\Delta S_\text{N} = \frac{k_\text{B}\Omega\phi_2}{n_2}\ln(\delta_{\text{max},2}). \tag{11}$$

Similarly, $\Delta S_\text{M}$ (see Figure 3 bottom) can be expressed as

$$\Delta S_\text{M} = \frac{k_\text{B}\Omega\phi_2}{n_2}\ln(\delta_{\text{max},2}). \tag{12}$$

Consequently, it is found that $\Delta S_\text{N}$ and $\Delta S_\text{M}$ are the same within the lattice theory, even though the proteins cannot take positions occupied by the polymers. That is, the structural stability of the protein in the aqueous polymer solution does not change within the theory. This result is somewhat exaggerated, but it agrees well with our experimental result from ultraviolet-visible absorption spectroscopy of hemoglobin in an aqueous starch solution (not shown). The result also roughly agrees with the thermal stabilities of proteins in an aqueous hyaluronic acid solution [15].

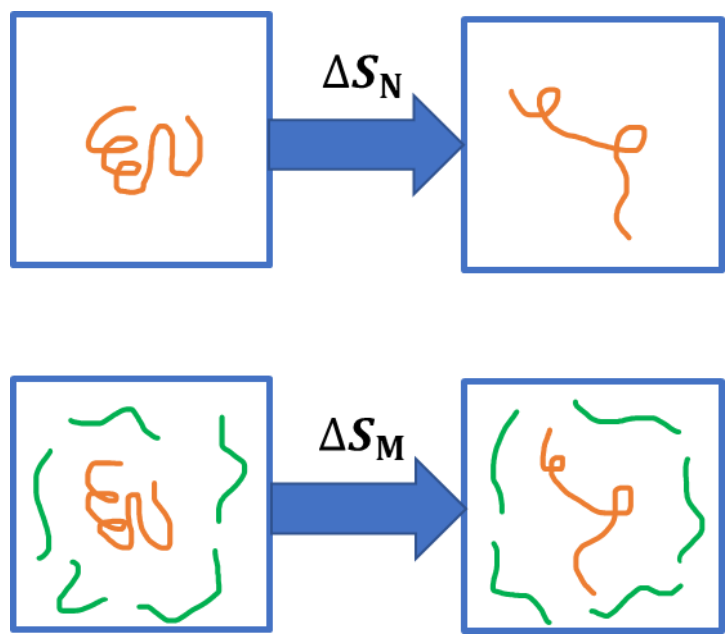

Figure 3: Illustration of the denaturation (unfolding). In the upper figure, the protein (orange) is denatured in aqueous solution, where the related entropy change is expressed as $\Delta S_N$. In the bottom figure, the protein is denatured in aqueous polymer solution, where the related entropy change is expressed as $\Delta S_M$. The polymers are drawn in green color. We found from the lattice theory that $\Delta S_N$ and $\Delta S_M$ are the same, even though the aqueous polymer solution is more crowded than the aqueous solution. Similar results have been also obtained from our spectroscopic measurements (not shown). It is expected that this property is helpful when the polymers are dosed to realize the NSSA effect.

In summary, we proposed and explained the ERAA effect and the NSSA effect. We focused on the repulsion between the normal cells and the drug, which has not been received much attention so far. We also focused on the NAO effect that the addition of the polymers enhances adsorptions of both the drugs and exosomes. We hope that our strategies will help the developments of the drugs from a theoretical physicochemical aspect and increase the delivery ratios.

**ACKNOWLEDGEMENTS**

We would like to thank M. Maebayashi, Y. Suzuki, Y. Uchida, and W. Furuta for useful discussion. We appreciate the experimental supports from M. Maebayashi, Y. Suzuki, and W. Furuta. This research was partially supported by JSPS KAKENHI (20K05437).

**REFERENCES**

[1] S. Wilhelm, A. J. Tavares, Q. Dai, S. Ohta, J. Audet, H. F. Dvorak, and W. C. W. Chan, Nature Reviews Materials, **1** (2016) 16014.

[2] K. Amano, M. Iwaki, K. Hashimoto, K. Fukami, N. Nishi, O. Takahashi, and T. Sakka, Langmuir, **32** (2016) 11063.

[3] K. Amano, T. Ishihara, K. Hashimoto, N. Ishida, K. Fukami, N. Nishi, and T. Sakka, Journal of Physical Chemistry B, **122** (2018) 4592.

[4] S. Furukawa, K. Amano, T. Ishihara, K. Hashimoto, N. Nishi, H. Onishi, and T. Sakka, Chemical Physics Letters, **734** (2019) 136705.

[5] K. Hashimoto, K. Amano, N. Nishi, and T. Sakka, Journal of Molecular Liquids, **294** (2019) 111584.

[6] K. Hashimoto, K. Amano, N. Nishi, and T. Sakka, Chemical Physics Letters, **754** (2020) 137666.

[7] J. C. Crocker, J. A. Matteo, A. D. Dinsmore, and A. G. Yodh, Physical Review Letters, **82** (1999) 4352.

[8] S. Ji and J.Y. Walz, Langmuir, **29** (2013) 15159.

[9] K. Amano, Y. Liang, K. Miyazawa, K. Kobayashi, K. Hashimoto, K. Fukami, N. Nishi, T. Sakka, H. Onishi, and T. Fukuma, Physical Chemistry Chemical Physics, **18** (2016) 15534.

[10] B. J. Lele and R. D. Tilton, Langmuir, **35** (2019) 15937.

[11] J. P. Hansen and I. R. McDonald, *Theory of simple liquids*, 3rd edition, Academic Press,

London, (2006).

[12] K. Amano, S. Furukawa, R. Ishii, K. Hashimoto, N. Nishi, and T. Sakka, Scientific Reports of the Faculty of Agriculture, Meijo University, **57** (2021) 9.

[13] M. C. Dixon, Journal of Biomolecular Techniques, **19** (2008) 151.

[14] F. Tanaka, *Koubunshi no Butsurigaku*, 4th edition, Shokabo, Tokyo, (2008).

[15] J. J. Water, M. M. Schack, A. Velazquez-Campoy, M. J. Maltesen, M. van de Weert, L. Jorgensen, European Journal of Pharmaceutics and Biopharmaceutics, **88** (2014) 325.